\documentclass[a4paper,11pt]{article}
\pdfoutput=1 

\usepackage{jheppub} 

\makeatletter
\gdef\@fpheader{ }
\makeatother

\usepackage{subcaption}
\renewcommand{\H}{{\mathcal{H}}}
\newcommand{\rrangle}{\rangle\!\rangle}

\title{SpaceTime from Hilbert Space: Decompositions of Hilbert Space as Instances of Time}

\author[a,b]{Mahdiyar Noorbala}

\affiliation[a]{Department of Physics, University of Tehran, Iran. P.O. Box 14395-547}
\affiliation[b]{School of Astronomy, Institute for Research in Fundamental Sciences (IPM), Tehran, Iran. P.O. Box 19395-5531}

\emailAdd{mnoorbala@ut.ac.ir}

\abstract{There has been recent interest in identifying entanglement as the fundamental concept from which space may emerge.  We note that the particular way that a Hilbert space is decomposed into tensor factors is important in what the resulting geometry looks like.  We then propose that time may be regarded as a variable that parameterizes a family of such decompositions, thus giving rise to a family of spatial geometries.  As a proof of concept, this idea is demonstrated in two toy models based on Kitaev's toric code, which feature a dynamical change of dimension and topology.}

\begin{document} 
\maketitle
\flushbottom

\section{Introduction} \label{sec:intro}

The idea that spacetime is an emergent notion has long been discussed in the physics literature, and there has been attempts to obtain Einstein's equations from such considerations \cite{J,V}.  Recently, there has been increasing motivation and overwhelming evidence that quantum entanglement may be the fundamental concept from which space emerges as a secondary construct \cite{vR,MS}.  An explicit such construction is recently proposed \cite{CCM}, in which the mutual information between various factors of an underlying Hilbert space is used to define a metric (distance) between these factors.  Although this and other such procedures can define space as an emergent concept from a purely quantum mechanical construction, they still have to deal with time as a fundamental notion to begin with.  This should be done by introducing an extra ingredient, namely a Hamiltonian, to generate the time evolution.  In this paper we propose that time, too, can be considered as an emergent concept arising from the same underlying Hilbert space without introducing a Hamiltonian.

As mentioned above, it is possible to define a metric on the factors $\H_p$ of a Hilbert space $\H=\bigotimes_p\H_p$.  These factors are considered as regions of space (or more accurately, points of the emergent metric space), but from the viewpoint of the underlying Hilbert space they are merely factors whose tensor product reproduces the entire space.  This factorization/decomposition is not unique in two ways.  (i) Some of the $\H_p$ may be further factored into smaller factors, or may be combined to form larger factors.  For example, consider
\begin{equation}
\H = \bigotimes_1^{10} \H_p = \H'_1 \otimes \H'_2 \otimes \H'_3,
\end{equation}
where $\H'_1 = \H_1 \otimes \H_2 \otimes \H_5 \otimes \H_9$, $\H'_2 = \H_3 \otimes \H_8 \otimes \H_{10}$, and $\H'_3 = \H_4 \otimes \H_6 \otimes \H_7$.  We call $\{\H_p\}$ and $\{\H'_p\}$ two \textbf{decompositions} for the same factorization $\H = \bigotimes_1^{10} \H_p$ of the Hilbert space.\footnote{\label{ft:decomp} $\{\H_p\}$ and $\{\H'_p\}$ are actually partitions that cover the entire space.  We do not insist that a decomposition must partition the Hilbert space; all we need is $\bigotimes_p\H_p \subseteq \H$ and $\H_p \cap \H_q = \{0\}$ for all $p\neq q$.  So in our constructions $\{ \H_1 \otimes \H_4, \H_5 \otimes \H_7 \}$ may well be considered as a decomposition.}  Different decompositions lead to different numbers of points in the metric space (e.g., $\{\H_p\}$ gives ten points while $\{\H'_p\}$ gives three points), and hence to different metrics and spatial geometries.  (ii) The two decompositions we mentioned can be obtained from each other by floating some factors around.  But it is also possible that factors of one factorization are non-trivially entangled with those of another.  An example is a two-spin system with basis vectors $\{ |00\rangle, |01\rangle, |10\rangle, |11\rangle \}$.  The standard factorization is manifest here: $\H = \H_{0,1} \otimes \H_{0,1}$, where $\H_{0,1}$ is a two-dimensional Hilbert space spanned by $|0\rangle$ and $|1\rangle$.  But we can also define
\begin{equation}\label{+-vs01}
\begin{aligned}
|++\rangle &:= |00\rangle, \\
|+-\rangle &:= \frac{1}{\sqrt2} (|01\rangle - |10\rangle), \\
|-+\rangle &:= \frac{1}{\sqrt2} (|01\rangle + |10\rangle), \\
|--\rangle &:= |11\rangle.
\end{aligned}
\end{equation}
The labeling of the LHS vectors suggests the factorization $\H = \H_{+,-} \otimes \H_{+,-}$, where $\H_{+,-}$ is another two-dimensional Hilbert space spanned by $|+\rangle$ and $|-\rangle$.  We call $\H = \H_{0,1} \otimes \H_{0,1}$ and $\H = \H_{+,-} \otimes \H_{+,-}$ two different \textbf{factorizations} of the same Hilbert space.\footnote{A more common name for factorization is ``tensor product structure''.  They are further discussed in Appendix~\ref{app:factorization}.  A detailed mathematical framework can be found in Ref.~\cite{ZLL}.}  Note that there is no way to write $|\pm\rangle$ in terms of $|0\rangle$ and $|1\rangle$, so $\H_{0,1} \neq \H_{+,-}$.  It is clear that a given factorization can lead to many decompositions.  Two distinct factorizations of a Hilbert space yield different numbers of points for the emergent metric space with different metrics/geometries.  Again these points correspond to regions of space.  But unlike different decompositions of a given factorization, one cannot identify the regions in one factorization as a mere re-organization of the regions of another factorization.  In the sequel, we use the word ``decomposition'' for a decomposition as well as its implied factorization.

In general, and in the absence of extra structures, there is nothing special about a given decomposition.  So the resulting geometry depends on the choice of the decomposition; and there are many of them.  We suggest that \textit{time is perhaps a variable to parameterize a family of decompositions.}  So we get a family of evolving spatial geometries.  This still doesn't clarify which family one should choose to work with, so there is a lot of arbitrariness here.  But this is not worrisome, since we already know from relativity that there are many choices for time.  In fact, a major motivation for this proposal comes from the fact that different observers see the universe differently.  If there is no observer around, then arguments from canonical quantum gravity suggest that there is no evolution, $H=0$, as implied by the Wheeler-de Witt equation \cite{W}.  This is consistent with our prescription in which the state of the total system has no dynamics.  However, if we are to describe the universe as viewed by an observer, then it is important where they are in the universe (i.e., which factor(s) of the total Hilbert space of the universe they occupy) and how they see and interact with the rest of the universe.  It is conceivable that this information is encoded in the decomposition of the Hilbert space, so the choice of time is ultimately related to the choice of the observer.  Ideas along this line are also discussed in Ref.~\cite{N}.  An argument supporting this point of view can be drawn from the fact that the set of controllable observables that are accessible to an observer induces a factorization of the Hilbert space \cite{ZLL}.  In this view, one may roughly say that time is related to a change in the accessibility of the universe to an observer.  We are not going to speculate how exactly such a relation can be defined, rather simply assume a family of decompositions to work with.

Our goal in this paper is to illustrate this idea via simple toy models based on Kitaev's toric code \cite{K}, which is briefly reviewed in Appendix~\ref{app:toric}.  These are not realistic models and are solely presented as a proof of concept.  In Section~\ref{sec:spatial} we review the procedure of Ref.~\cite{CCM} for obtaining spatial geometries given a decomposition.  Then in Section~\ref{sec:toys} we use two families of decompositions and compute the evolution of the spatial geometry in ``time'' in the context of two toy models.  Finally, in Section~\ref{sec:dis} we summarize and discuss some issues with our proposal, as well as possible future directions.

\section{Spatial Geometries} \label{sec:spatial}

In this section we review the procedure of Ref.~\cite{CCM} for obtaining the spatial geometry associated with a given state $|\psi\rangle$ and a given decomposition $\bigotimes_p \H_p$ of Hilbert space.  

Consider the weighted graph obtained from assigning each factor in the decomposition to a vertex.  The weight between vertices $p\neq q$ is given by\footnote{We use $\ln$ for natural logarithm and $\log$ for logarithm in base 2.}
\begin{equation}\label{def:w}
w(p,q) = - \ln \frac{I(p,q)}{I_\text{max}(p,q)},
\end{equation}
where $I_\text{max}(p,q)$ is the maximum mutual information (achievable under any state, not just $|\psi\rangle$) between the two Hilbert spaces $\H_p$ and $\H_q$.  Since $\H_p \cap \H_q = \{0\}$ we have $I_\text{max}(p,q) = 2\min\{ \log(\dim\H_p), \log(\dim\H_q) \}$. Next define
\begin{equation}\label{def:d}
d (p,q) = \min_\text{all paths connecting $p$ to $q$} \left( \sum_\text{along the path} w \right),
\end{equation}
which simply computes the shortest distance from $p$ to $q$ on the weighted graph.\footnote{We use $d$ to denote what the authors of Ref.~\cite{CCM} call $\tilde d$.}  Clearly, $d$ satisfies the triangle inequality and provides a metric space (not to be confused with an inner-product space) on the graph.

The next step is to embed the vertices of the graph on a Riemannian manifold isometrically (i.e., such that $d(p,q)$ coincides with the distance on manifold between $p$ and $q$).  There are obviously infinitely many such manifolds.  But in general it is a very hard task to find a low dimensional manifold that is as smooth as possible.  One strategy is embedding in a flat Euclidean space \cite{M} and visualizing the embedded graph as a meshed manifold.  Of course, there are still infinitely many manifolds that go through the points.  Even worse, there are examples, as simple as four points on $S^1$, that have no isometric embedding in $\mathbb{R}^D$ for any $D$.  Nonetheless, we can obtain a fair approximation if we use multidimensional scaling (MDS) \cite{BG}, as follows.

Suppose that an isometric embedding in $\mathbb{R}^D$ is possible for the $N$ vertices of the graph.  Let $X_p^i$ be the coordinates of the $p$th vertex ($1\leq p\leq N$ and $1\leq i\leq D$).  Assume that the $X_p$s are centered at the origin, i.e., $\sum_p X_p^i = 0$ for all $i$.  Then the components 
\begin{equation}\label{def:B}
B_{pq} = \sum_{i=1}^D X_p^i X_q^i
\end{equation}
of the matrix $B=XX^T$ can be read from the distances $d$ via
\begin{equation}\label{Bd}
B_{pq} = -\frac12 \left( d(p,q)^2 - \frac1N \sum_{q'=1}^N d(p,q')^2 - \frac1N \sum_{p'=1}^N d(p',q)^2 + \frac{1}{N^2} \sum_{p',q'=1}^N d(p',q')^2 \right).
\end{equation}
It is clear that $B=XX^T$ is a positive semi-definite matrix.  But given a set of distances $d(p,q)$, Eq.~\eqref{Bd} may not yield a positive semi-definite matrix.  However, if Eq.~\eqref{Bd} does yield a positive semi-definite matrix, then one can write it as $B=XX^T$ and obtain the embedding coordinates $X_p^i$ for an embedding in $\mathbb{R}^N$.  So a necessary and sufficient condition for existence of a flat isometric embedding is that $B\geq0$.  The $\mathbb{R}^N$ embedding has $X_p^i = \sqrt{\lambda_i} v^{(i)}_p$, where $\lambda_i$ is the $i$th eigenvalue of $B$ and $v^{(i)}$ is the corresponding eigenvector in an orthonormal set.  In some applications some of the $\lambda_i$s are zero, so we can easily have a $D$-dimensional embedding, where $N-D$ is the number of vanishing eigenvalues.  Even when no flat isometric embedding exists (i.e., some of the $\lambda_i$s are negative), we can work with the non-negative eigenvalues of $B$ and the corresponding eigenvectors.  In fact, usually the first few of the largest eigenvalues give a good approximate picture.  For example, if we take the eigenvectors $v^{(i)}$ ($i=1,\cdots,D$) corresponding to the first $D$ positive eigenvalues, then the distance matrix reads
\begin{equation}
d_\text{MDS}(p,q) = \sqrt{\sum_{i=1}^D (X_p^i-X_q^i)^2}.
\end{equation}
Indeed we use MDS merely for illustrative purposes, since even when a perfect embedding in flat space is feasible, it serves only as a mesh for the manifold.

Finally we define the discrepancy parameter
\begin{equation}\label{def:epsilon}
\varepsilon = \frac{\| d-d_\text{MDS} \|}{\| d \|},
\end{equation}
where $\|d\| = \sqrt{\operatorname{tr} (d^\dag d)}$ is the Hilbert-Schmidt norm of the matrix $d$.  $\varepsilon$ gives the relative error of the distances $d_\text{MDS}$ found by MDS with respect to the original distances $d$.

\section{Toy Models} \label{sec:toys}

To illustrate how time can be viewed as a variable parameterizing a family of decompositions, we consider as our toy model Kitaev's toric code \cite{K} on a genus-1 $k\times k$ square lattice (we work in units of the lattice spacing).  A brief overview of the toric code is provided in Appendix~\ref{app:toric}.  In fact, what matters here is not the toric Hamiltonian, rather the Hilbert space of the spins.  In the next two subsections we study two different families of decompositions for the Hilbert space of the toric code.  In both cases the system is in the ground state $|\psi\rangle = |\xi_{00}\rangle$, so we work exclusively with the pure density matrix $\rho = |\psi\rangle\langle\psi|$.

\subsection{Model A} \label{ssec:A}

\begin{figure}[tbp]
\centering
\begin{subfigure}{.5\textwidth}
\centering
\includegraphics{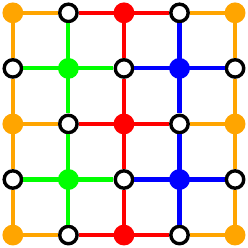} 
\caption{\label{fig:tilinga}}
\end{subfigure}~
\begin{subfigure}{.5\textwidth}
\centering
\includegraphics{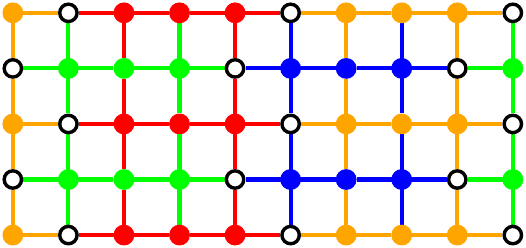} 
\caption{\label{fig:tilingb}}
\end{subfigure}
\begin{subfigure}{\textwidth}
\centering
\includegraphics{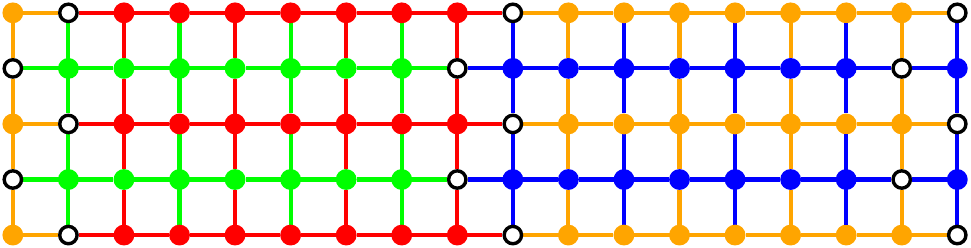} 
\caption{\label{fig:tilingc}}
\end{subfigure}
\caption{\label{fig:tiling} Decompositions (tilings) of the toric lattice (only a part of the lattice is shown).  Empty circles denote boundaries of factors, while solid ones join spins within a factor.  (a) Decomposition at $t=0$ as tiling with $m=1$ stars (+).  (b) Decomposition at $t=1$ as tiling with $m=2$ stars (++).  (c) Decomposition at $t=2$ as tiling with $m=4$ stars (++++).}
\end{figure}

In our first example, we use the standard factorization of the Hilbert space of the spin system,
\begin{equation} \label{standard-factorization}
\H = \bigotimes_i \H_i,
\end{equation}
where $i$ runs over all spins (links) on the lattice.  We label the decompositions by an integer $t$.  At $t=0$ the decomposition is equivalent to tiling the lattice with stars (+ shaped regions formed by four spins meeting at a vertex), as depicted in Fig.~\ref{fig:tilinga}.  At $t=1$ the decomposition is given by tiling the lattice with two neighboring stars; each factor (tile) looks like a ++ shape (see Fig.~\ref{fig:tilingb}).  Note that the arrangement here is such that the factors cover the whole lattice, so that our decompositions is a partition of the entire Hilbert space.  This is actually a coarse-graining, as we are enlarging the factors with increasing $t$.  We can move on and tile the lattice at time $t$ with factors each of the form ++\ldots+ and made up of $2^t$ stars.  Factors are arranged such that horizontal neighbors touch at only one vertex.  As we move vertically, factors are alternatively shifted by one lattice unit once to the right and once to the left (see Fig.~\ref{fig:tilingc} for $t=2$).  We demand that we have a $k\times k$ lattice with $k=2^{T+1}$ a power of 2, so that our tilings consistently cover the entire torus for $0\leq t \leq T$.

We now need to compute the mutual information between all pairs of factors of Hilbert space.  At $t=0$ we need the following facts about the mutual information $I(p,q)$ between two distinct stars $p$ and $q$ (see Table~\ref{tab:Iw} and Appendix~\ref{app:toric}):
\begin{equation}
I(p,q) =
\begin{cases}
1 & \text{$p$ and $q$ are diagonal neighbors,} \\
0 & \text{otherwise}.
\end{cases}
\end{equation}
We also note that for any pair of stars $p\neq q$, each Hilbert space has dimension $2^4$; hence $I_\text{max}(p,q) = 2\log 2^4 = 8$.  It then follows from Eqs.~\eqref{def:w} and \eqref{def:d} that $d(p,q)/\ln 8$ is equal to the $L^1$-norm (Manhattan distance) on the $45^\circ$-rotated periodic lattice.

\begin{figure}[tbp]
\centering
\begin{subfigure}{.24\textwidth}
\centering
\includegraphics[width=\textwidth]{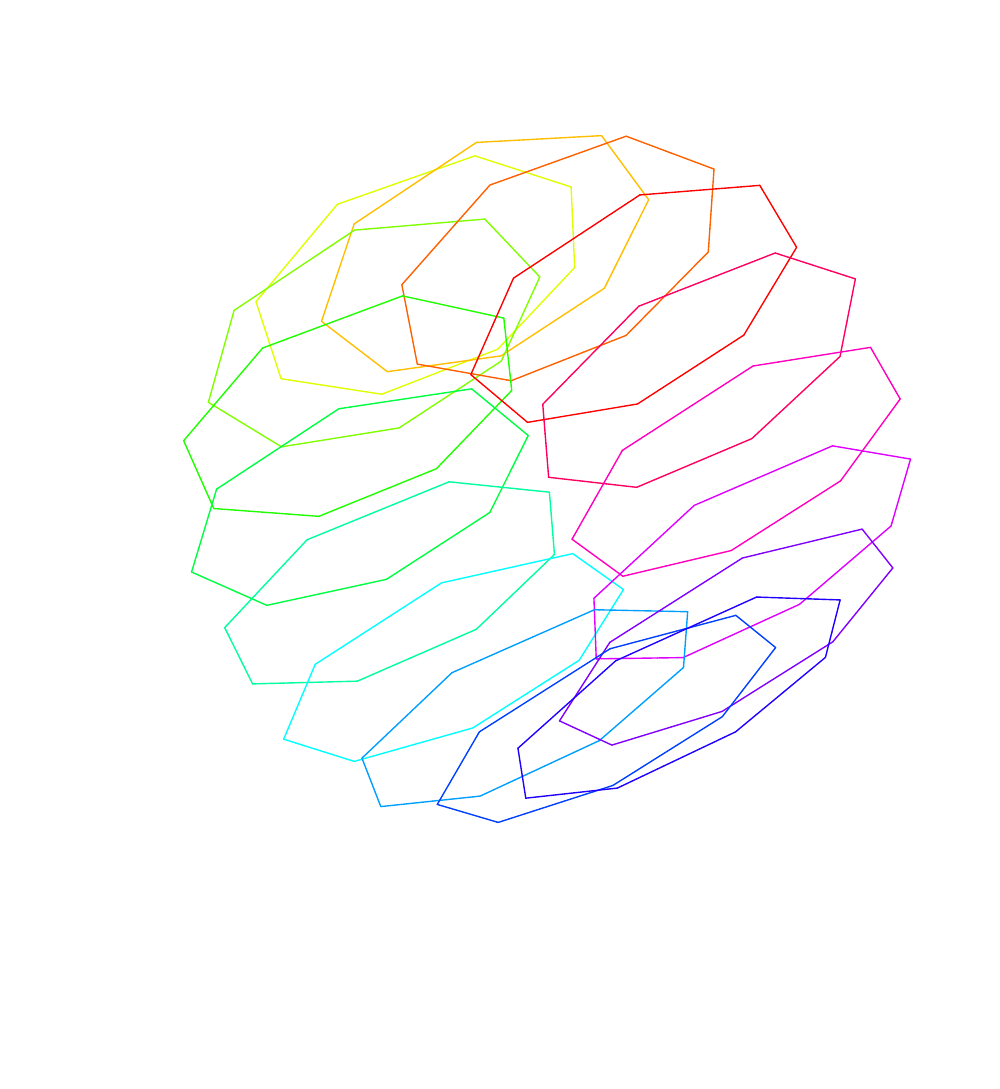}
\caption{\label{fig:t=0}}
\end{subfigure}
\begin{subfigure}{.24\textwidth}
\centering
\includegraphics[width=\textwidth]{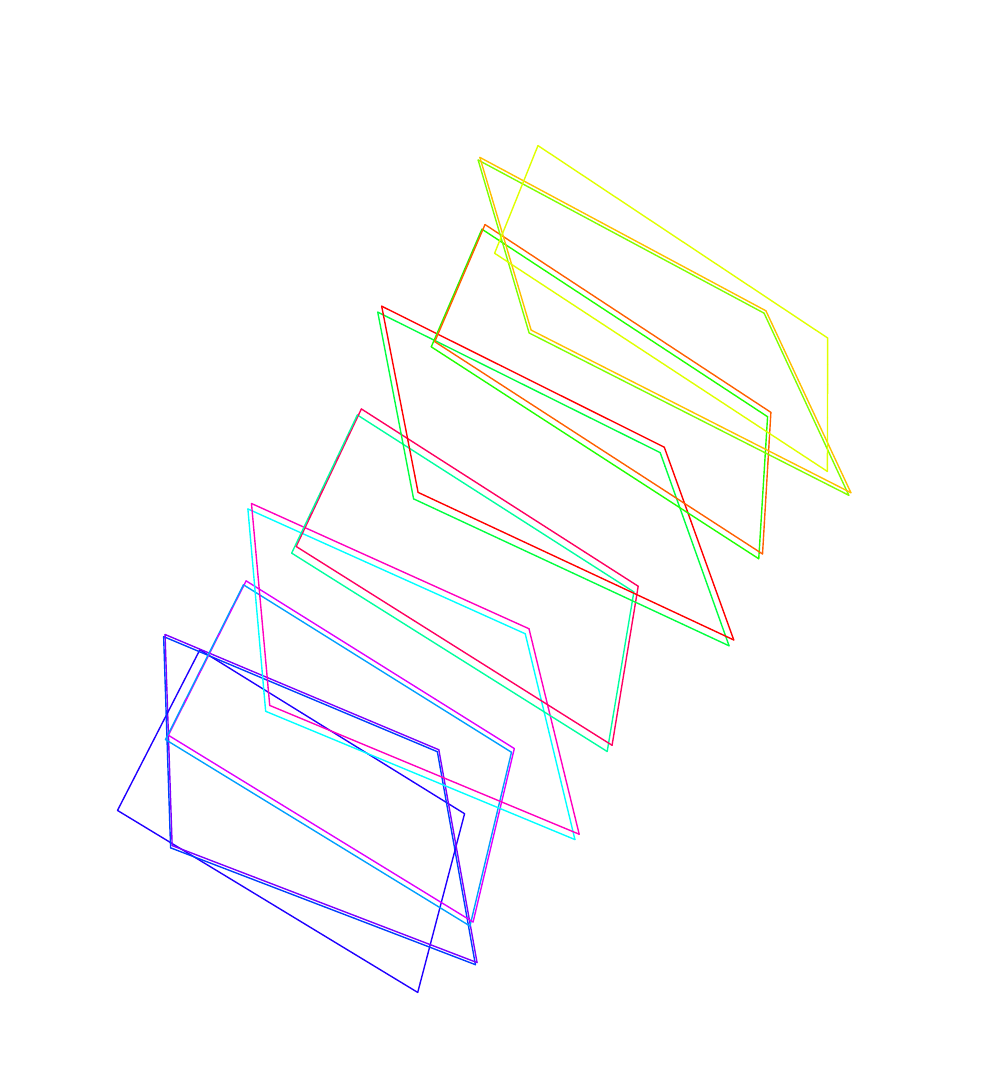}
\caption{\label{fig:t=1}}
\end{subfigure}
\begin{subfigure}{.24\textwidth}
\centering
\includegraphics[width=\textwidth]{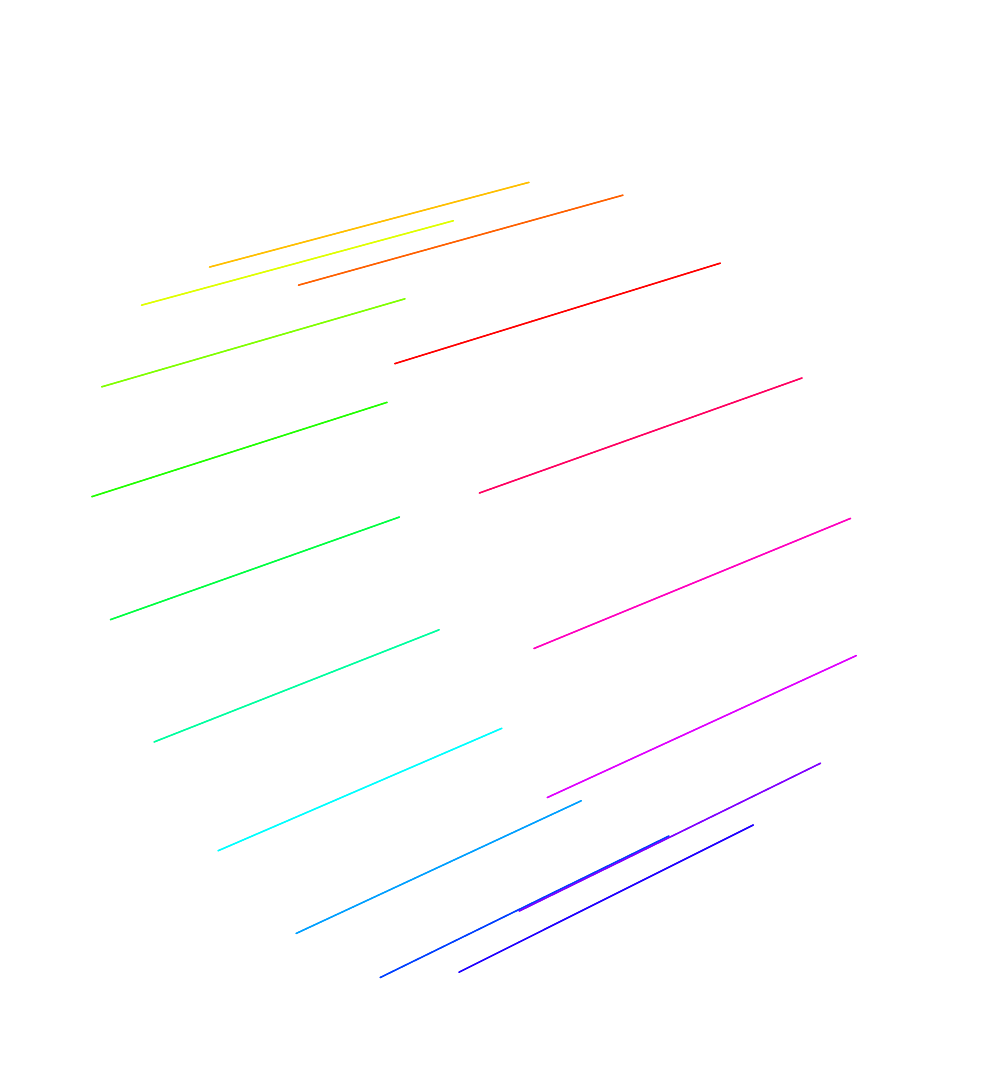}
\caption{\label{fig:t=2}}
\end{subfigure}
\begin{subfigure}{.24\textwidth}
\centering
\includegraphics[width=\textwidth]{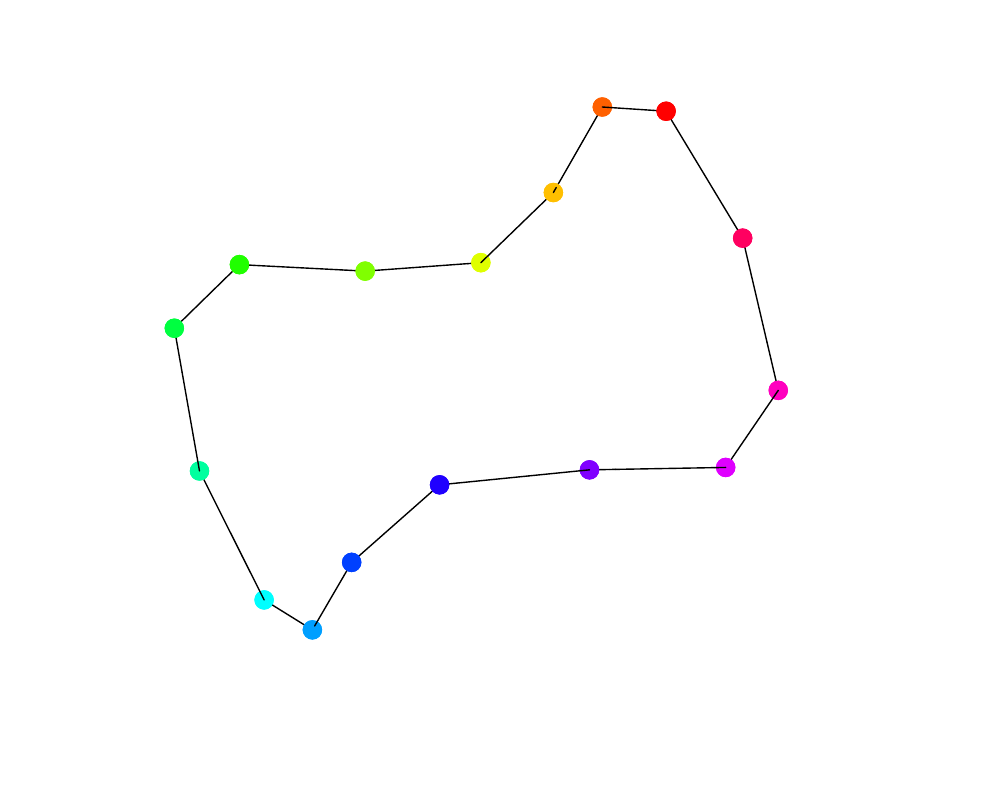}
\caption{\label{fig:t=3}}
\end{subfigure}
\caption{\label{fig:MDSgeometries} Evolution of spatial geometries in Model A constructed by MDS ($k=16$).  Adjacent factors in the same row of the lattice are connected (the color of the connecting line varies from row to row).  (a), (b), (c), and (d) show $t=0,1,2,3$ corresponding to $m=1,2,4,8$, respectively.  In (d) factors in neighboring columns are connected by a black line.}
\end{figure}

It remains to apply MDS to obtain an embedding for the spatial geometry in a flat space.  We have done this numerically for a $16\times16$ lattice.  The eigenvalues of the matrix $B$ defined by Eq.~\eqref{def:B} are $\{ 496_4, 219_4, \ldots, -219_4 \}$, where subscripts denote degeneracy.  Since there are negative eigenvalues, there is no flat isometric embedding.  Nevertheless, we can work with three of the dominant eigenvalues $\lambda_i$ ($i=1,2,3$) and use the corresponding eigenvectors $v^{(i)}$ to obtain coordinates $X^i_p = \sqrt{\lambda_i} v^{(i)}_p$ for an embedding in $\mathbb{R}^3$.  The result is shown in Fig.~\ref{fig:t=0}, which is pretty similar to a torus.  This is not surprising as $I$ and $d$ are indeed related to the usual distance on the toric lattice.  However, we emphasize that the actual metric space described by $d$ is not a torus, as the discrepancy parameter defined in Eq.~\eqref{def:epsilon} is found to be $\varepsilon = 0.28 \neq 0$.  What we have is a bunch of discrete points that at this level of MDS approximation look like points on a torus.  

\begin{figure}[tbp]
\centering
\begin{subfigure}{\textwidth}
\centering
\includegraphics{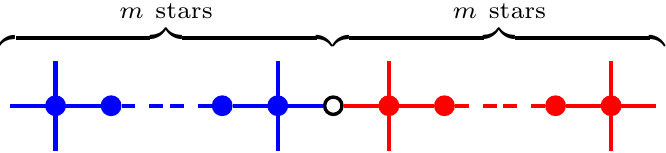} 
\caption{\label{fig:+m+a}}
\end{subfigure}
\begin{subfigure}{\textwidth}
\centering
\includegraphics{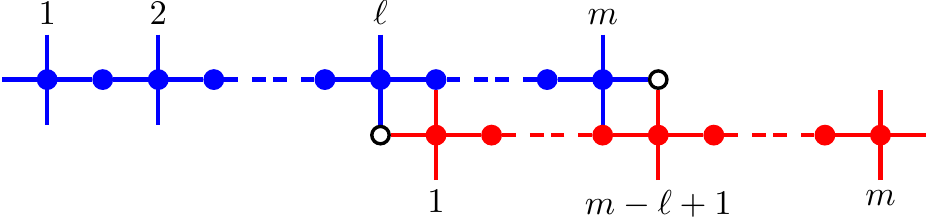} 
\caption{\label{fig:+m+b}}
\end{subfigure}
\begin{subfigure}{\textwidth}
\centering
\includegraphics{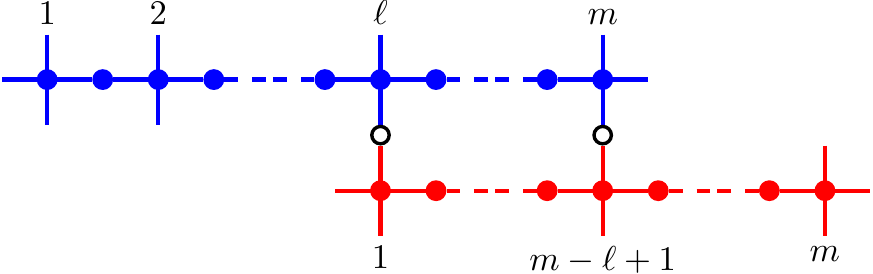} 
\caption{\label{fig:+m+c}}
\end{subfigure}
\caption{\label{fig:+m+} Relative position of neighboring factors.  (a) Horizontal neighbors.  (b) Neighbors on consecutive rows.  (c) Neighbors separated by a row.}
\end{figure}

As we move on to $t>0$, we need to compute the mutual information between more complicated factors in the Hilbert space.  The general result is summarized in Table~\ref{tab:Iw} and is as follows (the sketch of the calculation is presented in Appendix~\ref{app:toric}).  Let $m=2^t$ be the number of stars in any one factor of the decomposition at time $0\leq t<T$.  This factor of the Hilbert space has dimension $2^{4m}$, so $I_\text{max}(p,q)=8m$ for all $p\neq q$.  Any two such factors that have no common vertex have zero mutual information.  Also horizontal neighbors (which touch at one vertex as in Fig.~\ref{fig:+m+a}) have vanishing mutual information.  Now suppose that we have two factors lying on consecutive rows of the lattice, which are shifted by $2\ell-1$ lattice units (Fig.~\ref{fig:+m+b}).  Then the mutual information between these two factors is $2m-2\ell+1$.  The only remaining situation is when we have two factors that are separated by a row of the lattice, which are shifted by $2\ell-2$ lattice units (Fig.~\ref{fig:+m+c}).  In this case the mutual information between these two factors is $m-\ell$.

\begin{table}[tbp]
\centering
\begin{tabular}{||c||c|c||c|c||}
\hline\hline
& \multicolumn{2}{|c||}{$m<\frac{k}{2}$} & \multicolumn{2}{|c||}{$m=\frac{k}{2}$} \\
\hline\hline
Neighbor Type & $I(p,q)$ & $w(p,q)$ & $I(p,q)$ & $w(p,q)$ \\
\hline\hline
Fig.~\ref{fig:+m+a} & $0$ & $\infty$ & $0$ & $\infty$ \\
\hline
Fig.~\ref{fig:+m+b} & $2m-2\ell+1$ & $\ln\frac{8m}{2m-2\ell+1}$ & $2m+1$ & $\ln\frac{8m}{2m+1}$ \\
\hline
Fig.~\ref{fig:+m+c} & $m-\ell$ & $\ln\frac{8m}{m-\ell}$ & $m+1$ & $\ln\frac{8m}{m+1}$ \\
\hline\hline
\end{tabular}
\caption{\label{tab:Iw} Mutual information $I$ and weight factor of graph edges $w$ between different types of neighboring factors.}
\end{table}

\begin{figure}[tbp]
\centering
\begin{subfigure}{.21\textwidth}
\centering
\includegraphics[width=\textwidth]{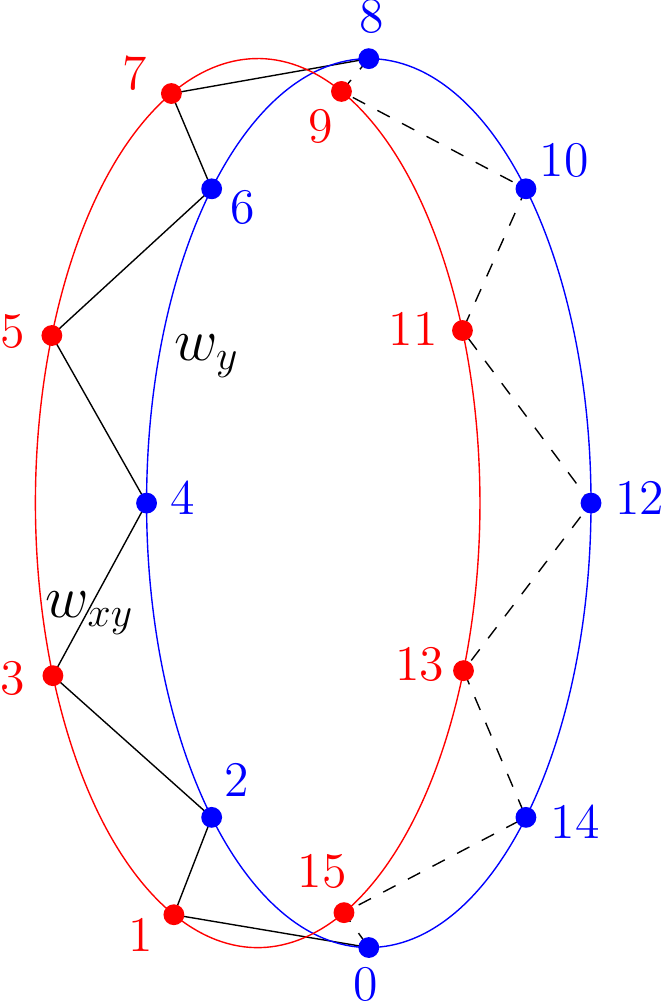} 
\caption{\label{fig:k=2m}}
\end{subfigure}~
\begin{subfigure}{.6\textwidth}
\centering
\includegraphics[width=\textwidth]{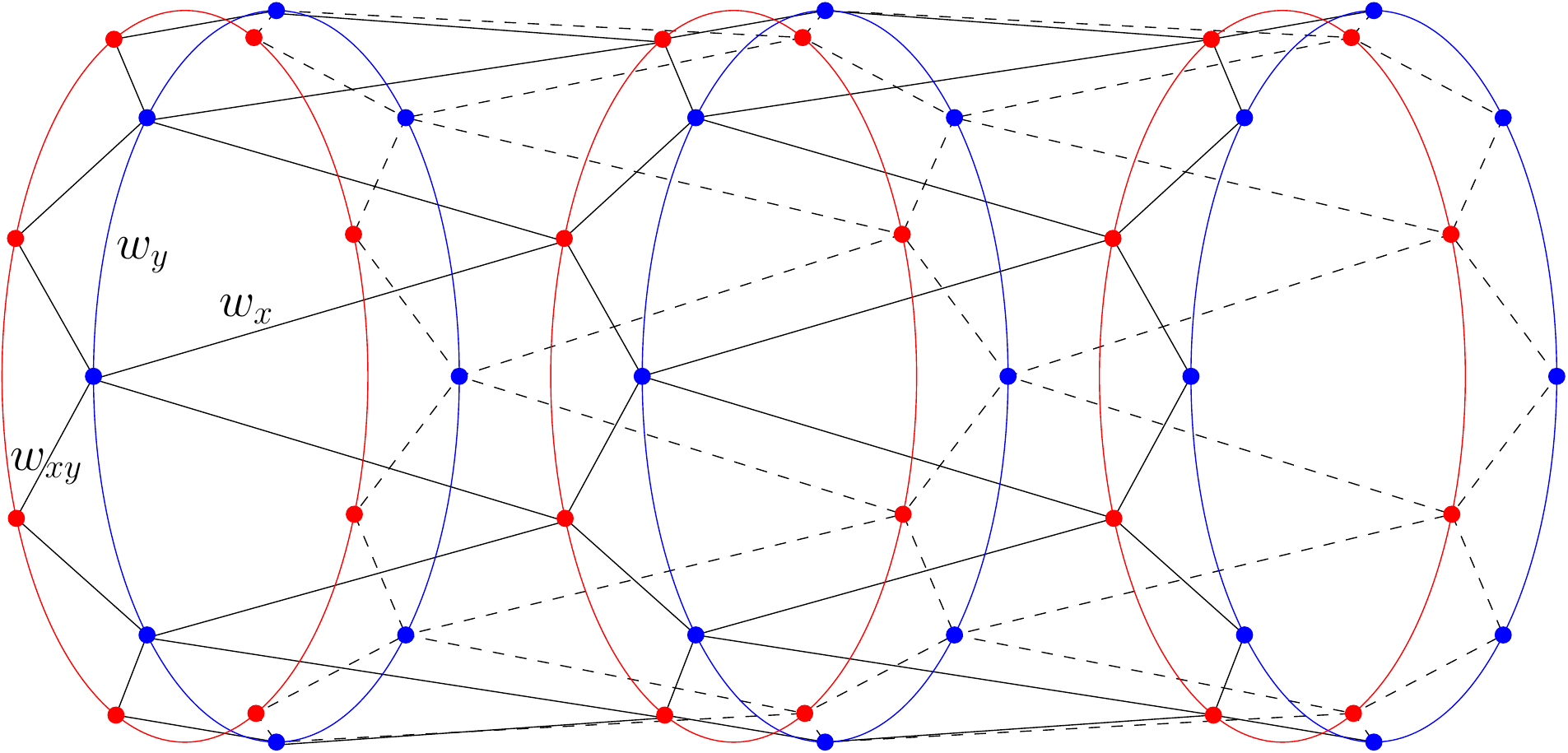} 
\caption{\label{fig:k>2m}}
\end{subfigure}
\caption{\label{fig:geometries} Graph of factors in Model A.  Each factor is represented by a vertex and distances are to be computed along the depicted edges with the given weights. (a) $k=8$ and $m=4$.  (b) $k=8$ and $m=2$; rightmost and leftmost circles of the same color are identified.}
\end{figure}

The above results apply to factors that are too far away to feel the topology of the torus.  At $t=T$ we have ring-like factors comprised of $m=k/2$ stars which encircle the torus.  In this case, there are no horizontal neighbors, nor is there any shift between vertical neighbors.  Factors lying on consecutive rows have mutual information $2m+1$, while those that are separated by a single row have mutual information $m+1$.  It follows that at $t=T$ the graph of factors has the shape of Fig.~\ref{fig:k=2m} with weights
\begin{equation}
w_y = \ln \frac{8m}{m+1} = \ln \frac{8k}{k+2}, \qquad w_{xy} = \ln \frac{8m}{2m+1} = \ln \frac{4k}{k+1}.
\end{equation}
Therefore, since $w_y<2w_{xy}$,
\begin{equation}
d(p,p') = \lfloor \frac{\Delta p}{2} \rfloor \times w_y + (\Delta p \mod 2) \times w_{xy},
\end{equation}
where
$0\leq p,p'<k$ are row numbers of the two ring-like factors, and $\Delta p$ is the distance between them noting the periodicity, i.e.,
\begin{equation}\label{AbsMod}
\Delta p = \min \{ |p-p'|, k-|p-p'| \}.
\end{equation}
The output of the MDS algorithm for this graph is shown in Fig.~\ref{fig:t=3}.  The spectrum of the matrix $B$ is $\{ 133_2, 16_2, \ldots, -33_2 \}$ and the discrepancy parameter is $\varepsilon = 0.19$.

Let us elaborate on the geometry at $t=T$.  If we consider only even $p$s, then we have $\frac k2$ points evenly distributed on a circle with circumference $\frac{k}{2} \ln\frac{8k}{k+2}$ (the blue points in Fig.~\ref{fig:k=2m}).  The odd $p$s also form a similar circle (the red points in Fig.~\ref{fig:k=2m}).  The entire geometry can be thought of as the union of the two circles as shown in Fig.~\ref{fig:k=2m}.  The allowed paths are those along the depicted lines.  Of course, this is not a geometry in the usual sense.  In particular, it is not a Riemannian manifold.  But one can imagine working with half of the factors (e.g., the even ones).\footnote{Technically, if we want to consider time evolution from $t=0$, then we need stars in both even and odd rows.  Otherwise when $t=0$ ($m=1$) and in the absence of odd stars, the even stars have infinite distance from each other.}  Then, in the large $k$ limit, this is equivalent to a 1D world that is fundamentally discrete with a ``Planck length'' of $\ln8$ (in lattice spacing units), but which at low energies appears as a world with a smooth geometry $S^1$.

This is by itself very interesting.  We have found a family of decompositions that interpolate between a 2D geometry ($S^1\times S^1$) and a 1D one ($S^1$).  It is an example of how geometry, as well as topology, can drastically change in the course of ``time'' as defined here.  Let us emphasize that we do not mean that $t$ is equivalent to the usual time generated by the Hamiltonian of the toric code.  We are merely using the toric code as an abstract Hilbert space for which a notion of time based on our proposal can be defined.  We do not imply that this describes the time in the toric model nor the time in the physical universe.  It is only an illustration of how the idea works.

Having discussed the initial and final geometries at $t=0,T$, we next turn attention to compute $d$ for the intermediate time steps $0<t<T$.  Using the information of Table~\ref{tab:Iw}, we find that the graph of factors has a structure similar to Fig.~\ref{fig:k>2m} (for smaller values of $m$ there are more rings but still the two sides are identified) with
\begin{equation}
w_x = \ln 8m, \qquad w_y = \ln \frac{8m}{m-1}, \qquad w_{xy} = \ln \frac{8m}{2m-1}.
\end{equation}
For a generic graph one can use the Floyd-Warshall algorithm to find the shortest path on a weighted graph.  But in our case we can write, with a little bit of reflection, an analytical result valid for $m>1$
\begin{equation}
d(p,q) = x w_x + (\Delta x - 2mx + x) w_{xy} + \theta(\Delta y - 2x) \frac{\Delta y - \Delta x + 2mx - 2x}{2} w_y,
\end{equation}
where $\theta$ is the Heaviside step function, $(\Delta x, \Delta y)$ is the coordinate difference between the factors $p$ and $q$ on the lattice (with periodicity taken into account as in Eq.~\eqref{AbsMod}), and $x=[\Delta x/2m]$ is the nearest integer to $\Delta x/2m$.

The MDS outputs for $t=1,2$ on a $16\times16$ lattice (corresponding to $m=2,4$) are presented in Fig.~\ref{fig:geometries}.  In both cases, we have kept only three eigenvectors of $B$ to work with.  For $t=2$, the spectrum of $B$ is $\{ 367_2, 230, 42, \ldots, -85_2 \}$ and the discrepancy parameter is $\varepsilon = 0.20$.  The resulting picture is Fig.~\ref{fig:t=2} in which the line segments are actually very squeezed rectangles.  For $t=1$, the eigenvalues of $B$ are $\{ 917_2, 470_2, \ldots, -302_2 \}$ and we get the same picture as $t=2$ with $\varepsilon = 0.20$ if we choose the first three eigenvalues.  However, if we pick the three eigenvalues $917,470,470$, we find Fig.~\ref{fig:t=1}, which is worse in terms of the discrepancy parameter $\varepsilon = 0.29$, but gives a better view of stretched rectangles and the four factors lying on them.

\subsection{Model B} \label{ssec:B}

As our second example, we use a different factorization for the Hilbert space of $2k^2$ spins of the toric model.  Here we do not define factors as regions of the lattice.  To describe our factorization, we use the orthonormal basis $|\xi_1,\xi_2,A_s,B_p\rrangle$ consisting of the eigenvectors of the toric Hamiltonian,\footnote{As we said before, we don't use the Hamiltonian to generate time evolution.  It is used solely to construct a non-standard factorization.} reviewed in Appendix~\ref{app:toric}.  We call $|\sigma^z_1, \ldots, \sigma^z_{2k^2} \rangle$ the standard (computational) basis, and call $|\xi_1,\xi_2,A_s,B_p\rrangle$ the non-standard (energy) basis.  As explained in Appendix~\ref{app:factorization}, the fact that $|\xi_1,\xi_2,A_s,B_p\rrangle$ are orthonormal vectors is sufficient to define the isomorphism
\begin{equation}
|\xi_1,\xi_2,A_1,\ldots,B_1,\ldots\rrangle \mapsto |\xi_1\rrangle \otimes |\xi_2\rrangle \otimes |A_1\rrangle \otimes \ldots \otimes |B_1\rrangle \otimes \ldots
\end{equation}
and interpret this as a new factorization of our Hilbert space, namely,
\begin{equation} \label{non-standard-factorization}
\H = \H_{+,-}^2 \otimes \H_{+,-}^{k^2-1} \otimes \H_{+,-}^{k^2-1},
\end{equation}
where the first two factors correspond to $\xi_{1,2}$, the next $k^2-1$ to $A_s$s, and the last $k^2-1$ to $B_p$s.

Having introduced a non-trivial factorization~\eqref{non-standard-factorization} that is different from the standard one~\eqref{standard-factorization}, we now employ a family of decompositions that go continuously from one to the other.  Define
\begin{equation}
|s_1, s_2, \ldots \rangle_t := U(t) |s_1, s_2, \ldots \rangle,
\end{equation}
where each $s_i$ is either $+$ or $-$, and $U(t)$ is a one-parameter family of unitary operators such that $U(0)=1$ and
\begin{equation}
U(T) = \sum_{s_i} |s_1, s_2, \ldots \rrangle \langle s_1, s_2, \ldots |.
\end{equation}
In other words, $U(t)$ takes the standard (computational) basis $|\cdot\rangle = |\cdot\rangle_0$ and rotates it continuously\footnote{We can actually choose $t$ to be a continuous as well as discrete parameter here (in Model A it was necessarily discrete).} to the non-standard (energy) basis $|\cdot\rrangle = |\cdot\rangle_T$.  We can take $U(t)=[U(T)]^{t/T}$.  At each instance of time $t$, $|s_1, s_2, \ldots \rangle_t$ serves to define a factorization of the Hilbert space consisting of $2k^2$ elementary factors.  At $t=0$ we use the decomposition shown in Fig.~\ref{fig:tilinga} into $k^2/2$ star-shaped factors.  At all subsequent times we use the same decomposition, bearing in mind that the star-shaped factors are now defined with respect to the labels of $|s_1, s_2, \ldots \rangle_t$, rather than the spin links of the $t=0$ case.  In order to do so, we actually need to map the labels in $|s_1, s_2, \ldots \rangle_t$ to the links on the lattice.  For $t=T$ the details of this map are irrelevant, as all factors are in pure states anyway.

It is clear that at $t=0$ we have the same torus-like geometry as in Model A (Fig.~\ref{fig:t=0}).  However, at $t=T$, all factors are in pure states and there is no mutual information between any pairs of factors.  It therefore follows that the geometry at $t=T$ consists of $k^2/2$ infinitely-separated points.  This is again an example of a dramatic change in geometry in the course of ``time'' as defined by a family of decompositions of Hilbert space.  Although we have not computed the spatial geometries for intermediate times $0<t<T$, it is plausible that as $t$ changes continuously, they interpolate between the initial and final geometry.  Therefore, we have a situation similar to an expanding universe, where a finite number of points are followed in time.

\section{Discussion} \label{sec:dis}

The Hilbert spaces of quantum mechanical systems are usually endowed with a decomposition into tensor factors.  This is often not just extra mathematical structure, but also extra physics.  The decomposition usually refers to physical entities and is interpreted as their position.  This is true in spin systems (like the toric code), as well as in quantum field theories where the notions of space and locality enter through the assignment of position labels to factors of the Hilbert space.  Our approach in this paper is to free the Hilbert space from such a preferred decomposition.  We then attribute the notion of time to a variable that parameterizes a family of decompositions of the Hilbert space.  Thus giving up a prescribed concept of space, we can introduce the notion of time.

Indeed, if the ideas advertised in Refs.~\cite{vR,MS} are to be taken seriously, then the concept of space (distance) has to be defined via entanglement between factors of the Hilbert space in each decomposition.  Here we have followed Ref.~\cite{CCM} in defining distance, which is based on using mutual information between factors of the Hilbert space.  For our purposes in this paper, this is a suitable setup to show how our idea works.  But in general, there may be other ways to define distance that are more suitable.  In particular, mutual information is not a genuine measure of quantum entanglement---it receives contributions from classical correlations as well.  One may work with other measures like quantum discord or concurrence.  Also, the procedure outlined in Ref.~\cite{CCM} involves Eq.~\eqref{def:d} to guarantee triangle inequality.  However, it is rather at odds with the spirit of the original ideas that relate distance to entanglement (the distance between two regions of space is thought to be directly related to the entanglement between the very two regions themselves, not other regions).  At any rate, the final result of this or other potential prescriptions is a spatial geometry, to which we can apply our proposal to obtain spacetime.  We emphasize again that we use the word ``geometry'' as an approximate---but not exact---synonym for Riemannian manifold.  This is an approximation since we have a finite number of points with given mutual distances, not a smooth topological space that is locally homeomorphic to $\mathbb{R}^n$.

Using the prescription of Ref.~\cite{CCM} or any other prescription that is based on the idea of ``space from entanglement'' \cite{vR,MS}, one still has to pick a preferred decompositions of the Hilbert space to define entanglement between its factors.  We have exploited this arbitrariness to define a family of such decompositions, define a spatial geometry for each member of the family, and interpret this set of geometries as an evolving space, hence spacetime.

We have seen that our approach can generate dynamical change in dimension and topology of the spatial geometry.  This is quite interesting and shows the richness of the spacetime geometries that can be obtained this way.  The formalism allows both discrete and continuous time steps, as is the case with Model A and B, respectively.  Of course, this is only a toy model and does not describe the physical universe.  To come up with a real-world model which describes a field theory on a curved (or even flat) spacetime would require a non-trivial Hilbert space and a non-trivial family of its decompositions, neither of which we claim to have presented here.

There are clearly some issues with our proposal that we present here as open problems.  First of all, what is the state of the total Hilbert space?  And then, what is the family of decompositions that is to be used for a given observer?  The first question can be asked in conventional physics too (as the problem of initial conditions), and perhaps has no simple answer.  Although we have suggested that the answer to the second question should depend on how the observer accesses, probes, and interacts with the universe, we have not been specific.  Indeed it seems that there exists a degeneracy between various choices of the ordered pair (state, decomposition), giving rise to the same geometry.  So maybe an equivalence class is to be introduced.  At any rate, we seem to require a set of plausibility conditions on the pair (state, decomposition) to have geometries for the real universe that are compatible with Lorentz invariance and general relativity at low energy.  For example, the authors of Ref.~\cite{CCM} suggest one such criterion via the idea of ``redundancy-constrained states''.  We do not speculate about these conditions here, but let us just say that if the state of the total Hilbert space is pure, then the holographic principle \cite{tH,S} implies that the spatial geometry must be compact (have no boundary).

Perhaps the most important missing ingredient in our proposal is how to glue the spatial geometries together to form a spacetime.  In particular, we have to specify the proper time between two factors belonging to consecutive decompositions, so as to know how the proper time changes as one passes through the spatial geometries.  This is also related to the arrow of time.  In its current form, our proposal allows reversal of time direction by simply reversing the order of the decompositions within a family.  To induce an intrinsic arrow of time, proper time may be defined to flip sign when the order of the decompositions change.  There are information-theoretic quantities that are asymmetric, so in principle such a definition of proper time should be possible.  We leave a detailed study of this to a future work.

\appendix

\section{Factorizations of Hilbert Space} \label{app:factorization}

In this appendix we elaborate on non-trivial factorizations of Hilbert space.  Below we show that any orthonormal basis of a Hilbert space induces a factorization (tensor product structure) of that Hilbert space.  The examples presented in Eqs.~\eqref{+-vs01} and \eqref{non-standard-factorization} are special cases of this fact.  We then clarify how to compute the entropy of a factor.

Let $\H$ be a finite-dimensional Hilbert space and let $|x_1,\ldots,x_n\rangle$ be an orthonormal basis of $\H$ with independent labels $x_i \in X_i = \{1,\ldots,d_i\}$ for $1\leq i\leq n$.  It may be the case that there is a maximally commuting set of observables whose eigenvalues $x_i$ label these vectors.  But in general, an arbitrary such labeling exists whenever $\dim\H = d_1\ldots d_n$.  (So there exists no non-trivial such basis for prime-dimensional Hilbert spaces.)  Clearly, we have
\begin{equation}
\langle x_1,\ldots,x_n | x'_1,\ldots,x'_n\rangle = \delta_{x_1x'_1} \ldots \delta_{x_nx'_n}.
\end{equation}
Now define
\begin{equation}
\H_i = \operatorname{span} \left\{ |x\rangle : x\in X_i \right\}
\end{equation}
to be a $d_i$-dimensional Hilbert space and consider the map $f:\H\to\H_1\otimes\ldots\otimes\H_n$
that acts like
\begin{equation}
f \left( |x_1,\ldots,x_n\rangle \right) = |x_1\rangle \otimes \ldots \otimes |x_n\rangle 
\end{equation}
on the basis vectors and is linearly extended to the entire domain $\H$ of $f$.  This map preserves the inner product: For any $|u\rangle,|v\rangle\in\H$, we have $\langle u|v\rangle = \langle U|V \rangle$, where $|U\rangle = f(|u\rangle)$ and $|V\rangle = f(|v\rangle)$.  Thus $f$ is an isomorphism between the Hilbert spaces $\H$ and $\H_1\otimes\ldots\otimes\H_n$.  We conclude that any basis of $\H$ that consists of $n$ labels induces a tensor product structure (a factorization) of $\H$ with $n$ factors, as well as a basis for each $\H_i$.  Conversely, it is obvious that given a factorization of $\H$, together with a basis for each of its factors, one can write a natural basis for the total Hilbert space $\H$.

The above discussion can be expressed in the mathematical framework developed in Ref.~\cite{ZLL} for tensor product structures on a Hilbert space.  The subalgebra ${\cal A}_i$ in that language corresponds to the algebra of operators that leave all labels except $x_i$ unchanged (isomorphic to $d_i\times d_i$ matrices).

Subsystems, for which quantities like entropy are calculated, correspond to certain factors in a preferred factorization.  When we have a family of factorizations, computation of entropy must be done with care.  Suppose each of the decompositions is induced from a basis $|x_1,\ldots,x_n\rangle_t$, as mentioned above.  So at each $t$
\begin{equation}
\H^t_i = \operatorname{span} \left\{ |x\rangle_t : x\in X_i \right\},
\end{equation}
and the isomorphism $\H=\bigotimes_i \H^t_i$ is realized by mapping $|x_1,\ldots,x_n\rangle_t$ to $|x_1\rangle_t \otimes \ldots \otimes |x_n\rangle_t$.\footnote{In a more accurate notation this would read $|x_1\rangle_t \otimes_t \ldots \otimes_t |x_n\rangle_t$ since the tensor product structures are different at different $t$.}

Let $I_t\subseteq\{1,\ldots,n\}$ be a set of factors in $\bigotimes_i \H^t_i$ for which we want to compute entropy.  The state of the total system is $\rho$ and is not changing.  However, the very meaning of a tensor factor is $t$-dependent, even when $I_t$ is $t$-independent.  So the density matrix of the factor (or subsystem) $I_t$ is given by
\begin{equation}
\rho_{I_t} = \operatorname{Tr}_{\bar{I_t}} \rho = \sum_{x_{\bar{I_t}}} \langle x_{\bar{I_t}} | \rho | x_{\bar{I_t}} \rangle,
\end{equation}
where $x_{\bar{I_t}}$ means all $x_i$ for which $i\notin I_t$, and $|x_{\bar{I_t}}\rangle = \bigotimes_{i\notin I_t} |x_i\rangle_t$ (note that here $\otimes$ means $\otimes_t$).  The entropy of this factor is
\begin{equation}
S(I_t) = -\operatorname{Tr}_{I_t} \left[ \rho_{I_t} \log \rho_{I_t} \right] = -\sum_{x_{I_t}} \langle x_{I_t} | \rho_{I_t} \log \rho_{I_t} | x_{I_t} \rangle,
\end{equation}
with a similarly defined notation.

\section{The Toric Code} \label{app:toric}

To be self-contained, in this appendix we briefly review basic facts about Kitaev's toric code \cite{K} that we have used in this paper.  Consider a $k\times k$ square lattice with periodic boundary conditions: the sites on the rightmost column are linked to those in the leftmost one, and those on the bottom row are linked to the top one.  So there are $k^2$ sites and $2k^2$ links altogether.  To each link we associate a spin (a quibt).

The four neighboring sites of a given site $s$ define a star and are denoted by ``$\text{star}(s)$''.  The four sides of a unit lattice cell define a plaquette denoted by ``$\text{boundary}(p)$'' and labeled by its center $p$, which is a point on the dual lattice.  There are $k^2$ star operators and $k^2$ plaquette operators defined as
\begin{equation}
A_s = \prod_{j\in\text{star}(s)} \sigma^x_j, \qquad B_p = \prod_{j\in\text{boundary}(p)} \sigma^z_j.
\end{equation}
They are Hermitian with eigenvalues $\pm1$ and commute with each other and with the Hamiltonian
\begin{equation}
H = -\sum_s A_s - \sum_p B_p.
\end{equation}
But since
\begin{equation}
\prod_s A_s = \prod_p B_p = 1,
\end{equation} 
they are short of being a complete set of commuting operators, i.e., the eigenvalues of $A_s$ and $B_p$ provide $2k^2-2$ independent numbers to label the eigenvectors of $H$.  The two additional labels required for a complete labeling of the eigenstates of $H$ are provided by the eigenvalues of
\begin{equation}
\xi_a = \prod_{j\in C_a} \sigma^z_j, \qquad a=1,2,
\end{equation}
where $C_{1,2}$ are two independent non-contractible cycles of the torus.  $\{\xi_1$, $\xi_2$, $A_1$, \ldots, $A_{k^2-1}$, $B_1$, \ldots, $B_{k^2-1}\}$ is thus a complete set of $2k^2$ independent commuting Hermitian operators, and hence we have the orthonormal basis $|\xi_1,\xi_2,A_s,B_p\rrangle$ labeled by the corresponding eigenvalues.  The notation $|\cdot\rrangle$ is employed to prevent confusion from the standard basis $|\sigma^z_1, \ldots, \sigma^z_{2k^2} \rangle$ where the $z$-components of spins are used for labeling.  Since all operators in $\{\xi_1$, $\xi_2$, $A_1$, \ldots, $A_{k^2-1}$, $B_1$, \ldots, $B_{k^2-1}\}$ commute with the Hamiltonian, $|\xi_1,\xi_2,A_s,B_p\rrangle$ are eigenvectors of $H$.  In particular, the four vectors with $A_s=B_p=1$ (namely, $|+++\ldots+\rrangle$, $|+-+\ldots+\rrangle$, $|-++\ldots+\rrangle$, and $|--+\ldots+\rrangle$)\footnote{In the literature, these are usually called $|\xi_{00}\rangle$, $|\xi_{01}\rangle$, $|\xi_{10}\rangle$, and $|\xi_{11}\rangle$, respectively.} span the ground state subspace.

We need to compute the von Neumann entropy for an arbitrary set $R$ of spins on the toric lattice, when the system is in the ground state $|\xi_{00}\rangle = |+\ldots+\rrangle$ of $H$.  It can be shown \cite{HIZ} that when $R$ is a connected set of spins then the entropy is given by
\begin{equation}\label{SofR}
S(R) = \Sigma_{R\bar R} - 1,
\end{equation}
where $\Sigma_{R\bar R}$ is the number of independent star operators that act on both $R$ and $\bar R$.  For example, consider the situation of Fig.~\ref{fig:+m+b}.  The left (blue) factor consists of $m$ stars.  To compute its entropy we notice that every vertex in this shape contributes 1 to $\Sigma_{R\bar R}$, except for the vertices that are centers of the $m$ stars.  Therefore, $\Sigma_{R\bar R} = 3m+1$.  Equation~\eqref{SofR} then gives $S(R)=3m$.  Now let $R$ be the union of the two (blue and red) factors in Fig.~\ref{fig:+m+b}.  In this case, the number of independent star operators acting on both $R$ and $\bar R$ is twice the previous number, minus the number of overlapping vertices, i.e., $\Sigma_{R\bar R} = 2(3m+1) - 2(m-\ell+1) = 4m+2\ell$.  Since this is also a connected set of spins, we can still use Eq.~\eqref{SofR} and obtain $S(R) = 4m+2\ell-1$.  Finally, we can put all of these together and compute the mutual information between the two (blue and red) factors:
\begin{equation}
\begin{aligned}
I(p,q) &= S(p) + S(q) - S(p,q) \\
&= (3m) + (3m) - (4m+2\ell-1) \\
&= 2m-2\ell+1.
\end{aligned}
\end{equation}
We obtain the other entries of Table~\ref{tab:Iw} by a similar counting.

\acknowledgments
I would like to thank Jahanfar Abouie and Vahid Karimipour for useful discussions.  I also appreciate the hospitality of the HECAP section at ICTP where this work was completed.  I acknowledge financial support from the research council of University of Tehran.

\end{document}